\documentclass[10pt,twocolumn,twoside]{article}

\usepackage[margin=2cm]{geometry}
\usepackage[utf8]{inputenc}
\usepackage[T1]{fontenc}
\usepackage{lmodern}

\usepackage{amsmath, amsfonts, amssymb}
\usepackage{graphicx}
\usepackage{booktabs} 
\usepackage{caption}
\usepackage{xcolor}
\usepackage{hyperref}
\usepackage{url}

\usepackage{authblk}

\newbox{\orcid}\sbox{\orcid}{\includegraphics[scale=0.1]{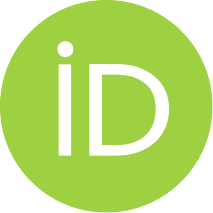}}

\usepackage[style=numeric,sorting=none,backend=biber]{biblatex}
\title{Shannon entropy and complex network community detection to study electoral coalition behaviour at municipal scale}
\author[1]{%
    \href{https://orcid.org/0009-0003-2239-9817}{\usebox{\orcid}}\hspace{1mm}Andrea Lo Sasso\thanks{\texttt{andrealosasso97@gmail.com}}%
}

\begin{document}

\maketitle

\section{Abstract}

Traditional electoral analyses often rely on aggregate sociopolitical indicators or on macroscopic models grounded in statistical physics that treat election data at national level; however, these models have rarely been applied to local-level voting data, where the analytical complexity is substantially higher. In this study, we propose a framework rooted in statistical physics and complex network theory to investigate the fine-grained architecture of voting behaviour at the local scale. Leveraging a granular dataset from the 2024 municipal and 2025 regional elections in Bari, Italy, we represent the urban electoral landscape as a complex network, in which polling sections are modelled as nodes connected by links that encode statistically significant correlations based on vote expression. For each coalition participating in the elections, community detection reveals geographically proximal clusters that transcend administrative boundaries. Furthermore, in order to quantify whether a coalition exhibits territorially homogeneous voting behaviour or, instead, is fragmented into distinct voting blocs, we adapt latent-ideology estimation and the information-theoretic Shannon entropy $H$ to the intra-coalition scale. Our results indicate that winning coalitions exhibit significantly higher territorial fragmentation, and these findings suggest that, at the local level, electoral success is not driven by the maintenance of geographically uniform consensus rather than by the capacity to aggregate diverse and non-homogeneous voting blocs. This scalable, data-driven framework moves beyond simple geographic accounting, providing a robust tool for uncovering the structural dynamics of political idea clustering and coalition fragmentation in complex urban environments.

\section{Introduction}

The study of political behaviour, long analysed within political science, sociology, and psychology, has witnessed a paradigm shift in the twenty-first century \cite{tufekci2014engineering, gsanger2024opinion} thanks to the explosion in the availability of high-resolution data regarding voting records. A new analytical framework incorporating the rigorous formalism of statistical mechanics and complex systems science has emerged, departing from traditional social science methodologies.  \cite{borghesi2012election, kaufman2022statistical}. The "socio-physics" posits that, much like interactions among particles produce collective physical phenomena, micro-level interactions among voters, modelled by social and media influences, generate emergent political outcomes that are analysable via quantitative techniques \cite{siegenfeld2020negative}.

Until now, the electoral phenomena have been modelled using the Ising framework, which assumes the presence of a binary choice, such as voting for Democrats or Republicans in US elections, or the Leave/Remain decision in the Brexit referendum, incorporating ferromagnetic or antiferromagnetic interaction terms alongside external fields representing mass media or candidate charisma \cite{gsanger2024opinion, kaufman2022statistical}. An extension of this approach is provided by the Potts model, which generalizes the Ising framework by introducing additional states (e.g., for Democrats, Republicans, and Independents in US elections), thereby enabling the representation of multiparty electoral systems, such as those in the UK or Canada \cite{kaufman2024social}. However, although these models successfully describe electoral systems at the national level by aggregating information across a vast number of measurements, they fail to provide information regarding the internal behaviour and the currents traversing a coalition, which is highly useful in political strategies and electoral list construction choices \cite{pal2025universal}.

Furthermore, because such analyses require the availability of enormous amounts of data, a microscopic description of electoral behaviour remains largely unexplored. This is particularly evident with regard to list-building strategies: specifically, whether it is more advantageous to assemble lists of candidates capable of mobilizing territorially homogeneous voting patterns, or to rely on "hub" candidates who can effectively aggregate heterogeneous voting blocs. Given that many electoral systems allow for the formation of coalitions supporting a single candidate (such as mayoral elections in municipal elections), these analyses represent a crucial aspect of quantitative political strategy; however, there is still no clear consensus on whether electoral success is more strongly associated with coalitions whose constituent parties homogeneously win the vote across the territory, or with coalitions dominated by a leading candidate who draws the majority of the coalition's votes. This distinction is already difficult to assess at the national level \cite{perc2017statistical}, and becomes even more complex to investigate at the local level. This is a phenomenon that cannot be adequately captured through a ferromagnetic–antiferromagnetic lens \cite{monroe2001paradigm}; rather, it unfolds through a configuration of relational interactions and/or through the structure of received votes, and therefore calls for new mathematical models capable of describing these dynamics effectively.

Electoral behaviour at the local level is then distinct from that observed in national elections. The literature shows that, at the local scale, interactions among geographically proximate voters are intense and influential, whereas this characteristic diminishes at the national scale, where distance attenuates social influence \cite{foladare1968effect}. These local interactions generate fluctuations in overall electoral behaviour that tend to be smoothed out in macroscopic analyses, while remaining markedly impactful at a microscopic scale, such as that of a single city \cite{auconi2024fluctuations}.

To overcome these limitations when transitioning from a macroscopic to a local analysis, numerous recent studies on such complex phenomena have increasingly adopted a complex-network approach, which allows for a deeper exploration of the collective relationship underlying electoral outcomes on a local scale. traditional statistical methods that focus on modelling overall electoral behaviour, such as the voting behaviour of individual candidates \cite{lyra2003generalized} or their socio-economic status \cite{hossain2017people}, a complex network analysis that explicitly models the vote interactions recorded in each electoral section becomes necessary \cite{coscia2012towards}. In fact, complex networks are a tool of statistical physics that allow for the modelling of complex phenomena, and they have already proven effective in a wide range of domains, including economics \cite{hidalgo2007product, battiston2012debtrank, bardoscia2021physics, lo2025exploring, bellantuono2022territorial, bellantuono2025network, lo2025exploring}, neuroscience \cite{amoroso2018multiplex, bellantuono2021predicting}, and social mobility \cite{batty2021london}, just to name a few.

The versatility of complex networks has also been employed in recent years to investigate electorate polarization through social media, demonstrating the potential of this methodology for sociopolitical analysis \cite{flamino2023political, di2024sampled, falkenberg2022growing}. Despite the work on political debate ideas, this same theory offers insights for studying the political-electoral scenario, useful for describing the dynamics of an election and providing food for thought for future electoral coalitions. Despite its potential to yield valuable insights for electoral strategy \cite{wagner2021affective, kam2017polarization, maza2022attempting}, this area of studies remains largely underexplored. Indeed, even when a coalition secures broad support, the same support is expected to be unevenly distributed across the territory \cite{wildgen1980spatial, barkan2006space}; consequently, understanding how to configure and coordinate coalition-level behaviour is instrumental for refining electoral strategies.

In this study, we propose a framework to analyze electoral behaviour, focusing on the city of Bari during the municipal elections of June 2024 and the regional elections of November 2025, in which different competing coalitions were recorded in the municipal and regional election. We construct a similarity network, in which nodes represent polling sections, and weighted links correspond to statistically significant correlations in voting patterns for each coalition competing in the electoral race. In addition, we extend the concept of latent ideology estimation to the intra-coalition level by applying singular value decomposition to the voting matrix of individual coalitions. This methodology reveals an emergent binarization within coalitions. By showing that the latent-ideology distribution of each coalition cannot be adequately approximated by a Gaussian, we introduce Shannon entropy as a principled measure to quantify the degree of non-uniformity in each coalition's ideological profile. In this way, it is possible to identify which coalitions exhibit relatively homogeneous spatial behaviour, indicated by low entropy values, and those characterized by pronounced internal fragmentation, as reflected by high entropy values. This allows to demonstrate which coalitions experience territorial fragmentation, compare these levels of fragmentation with the total votes received, and ultimately delineate the most effective strategic approach for coalitions at the local level.

The structure of the present work is as follows: the $Methods$ section details each stage of the model's development, from data collection to the validation of emergent communities and the state-of-art mathematical formulation of latent ideology and the Shannon definition of entropy. $Results$ section examines the principal findings of our approach: (a) the construction of a complex network of Bari's electoral sections, whose community structure reveals patterns of similarity among them; and (b) a set of streamlined procedures that leverage ideology analysis to reassess for each coalition how fragmented each coalition is by examining the entropy level. The $Discussion$ section considers the implications and added value of these results. Additional details regarding polling section, such as their location and other administrative details, are provided in the $Supplementary$ $Information$ \cite{lo_sasso_dataset_election}.

\section{Methods}
\subsection{Data collection}

In this section, we describe the procedure to build the complex network of each coalition on which our results are based. As outlined in the Introduction, the network construction process is applied consistently to each coalition. Our analysis separately examines the 2024 Bari municipal elections and the results of the 2025 Apulia regional elections specifically pertaining to the Bari constituency.
For what concerns mayoral election, data are downloaded from the election results on the municipality's statistical office website \footnote{\url{https://bari-risultati.palgpi.it/elezioni2024/comunali/Coalizioni/000054.html}}. Here, candidates have grouped into five coalitions. The mayoral candidates leading the coalition were: Mayor Candidate 1 (7 lists supporting, the winner coalition), Mayor Candidate 2 (10 lists supporting, the second-ranked coalition), Mayor Candidate 3 (6 lists supporting, the third-ranked coalition), Mayor Candidate 4 (1 list supporting, the fourth-ranked coalition) and Mayor Candidate 5 (2 lists supporting, the fifth-ranked coalition). As we show in the \emph{Complex network construction} section, due to statistical reasons we exclude from the following analysis the coalitions supporting Mayor Candidate 4 (1 lists) and Mayor Candidate 5 (2 lists).
In Table \ref{tab:statistica_datatset_mayoral}, we report the information regarding all candidates in the polling competition and the three main coalitions we analyze in the following sections.

Regarding the regional election, data are publicly available from Italian Ministry of the Interior's website (also known as $Eligendo$) that collects information on the results of election \footnote{\url{https://elezioni.interno.gov.it/risultati/20251123/regionali/elenchi/italia/16009009}}. Here, candidates have grouped into four coalitions: Regional President Candidate 1 (6 lists supporting, the winner coalition), Regional President Candidate 2 (4 lists supporting, the second-ranked coalition), Regional President Candidate 3 (1 lists supporting, the third-ranked coalition), Regional President Candidate 4 (1 list supporting, the fourth-ranked coalition). Similar to mayoral election, we exclude from the following analysis the coalitions supporting Regional President Candidate 3 (1 lists) and Regional President Candidate 4 (1 lists) for statistical reasons we will clarify in the next section. In Table \ref{tab:statistica_datatset_regional}, we report the information regarding all candidates in the polling competition and the two main coalitions we analyze in the following sections.

\begin{table}[t]
    \centering
    \resizebox{\linewidth}{!}{
    \renewcommand{\arraystretch}{1.25}
    \begin{tabular}{@{}lccc@{}}
        \toprule
        & \textbf{Candidates} & \textbf{Votes} & \textbf{Lists} \\ \midrule
        Full dataset & 850 & 118{.}414 & 26 \\
        Coalition 1  & 239 & 64{.}496  & 7  \\
        Coalition 2  & 319 & 30{.}306  & 10 \\
        Coalition 3  & 195 & 22{.}408  & 6  \\ \bottomrule
    \end{tabular}
    }
    \caption{Statistical summary of the dataset for mayoral election}
    \label{tab:statistica_datatset_mayoral}
\end{table}

\begin{table}[t]
    \centering
    \resizebox{\linewidth}{!}{
    \renewcommand{\arraystretch}{1.25}
    \begin{tabular}{@{}lccc@{}}
        \toprule
        & \textbf{Candidates} & \textbf{Votes} & \textbf{Lists} \\ \midrule
        Full dataset & 183 & 116{.}590 & 12 \\
        Coalition 1  & 96  & 81{.}866  & 6  \\
        Coalition 2  & 64  & 30{.}837  & 4  \\ \bottomrule
    \end{tabular}
    }
    \caption{Statistical summary of the dataset for regional election}
    \label{tab:statistica_datatset_regional}
\end{table}

\subsection{Complex network construction}
We collect all votes merging excel data grouped by list. We construct a matrix having the 345 section on the rows and the candidate on columns. In this way, each cell of the matrix represents the number of votes candidate $j$ received in section $i$. Finally, votes are normalized over the rows, in order to have a comparable behaviour over polling sections not relating it to the absolute number of preference of candidate records in a section.

We thus construct a weighted network in which nodes correspond to the electoral section, and the weighted link are the Pearson correlation of vote array recorded in two sections. Given the large number of potential weighted link in each network (345x(345-1) = 118680 potentially links per network), the adoption of a stringent significance threshold is necessary to account for multiple hypothesis testing \cite{foti2011nonparametric}, thereby limiting the inclusion of spurious connections and ensuring that only statistically robust and structurally meaningful links are retained. Thus, in this network two nodes are connected if the Pearson correlation between the sets of their vote expressed in is statistically significant, with $p-value \leq 10^{-6}$.

Furthermore, because the variance of the Pearson correlation coefficient depends strongly on sample size \cite{bates1996effects}, it is important limiting the analysis to correlations computed on sufficiently large vectors reduces estimation noise, in order to attenuate spurious high correlations caused by small sample fluctuations, and ensure that the inferred links are statistically robust. For this reason, in the mayoral election, we focus on vote recorded by the three largest coalitions and the two largest coalitions for regional election, as they have a enough number of votes making possible to study them by means of a statistical analyses.

As Table \ref{tab:statistica_datatset_mayoral}, the first coalition running for municipal election is referred to $C_{1}^{M}$ (centre-left coalition, having 240 candidates, and collecting 64.458 votes). The second coalition of municipal election is termed $C_{2}^{M}$ (right coalition, gathering 319 candidates, and 30.363 votes recorded), while the last coalition is named $C_{3}^{M}$ (left coalition, which records 195 candidates, and 22.408 votes). For what concerns the regional election (Table \ref{tab:statistica_datatset_regional}), we focus on vote recorded by the two largest coalitions, so defining $C_{1}^{R}$ (centre-left coalition, with 96 candidates and 81.866 votes) and $C_{2}^{R}$ (centre-right coalition, with 64 candidates and 30.837 votes).

\subsection{Community detection}
In network science, community detection on a weighted network identifies groups of nodes that are more strongly and densely connected to each other than to the rest of the network, taking edge weights into account. In this study, the community detection is performed using the Leiden algorithm (which is a modularity-based algorithm) \cite{traag2019louvain}, with the resolution $\gamma$ treated as a free parameter, and varying in [0.5, 1] with a 0.05 step, and the remaining algorithm parameters fixed to default ($\beta$ = 0.05, objective function = modularity). For each value of resolution, K = 100 algorithm runs are performed, each with a different pseudo-random number generator seed; we use majority voting to choose among the resulting partitions. The procedure is made more robust by using a stability criterion, that considers the similarity of different partitions $p_j$ with j=1,..,K, based on the average Normalized Mutual Information:

\begin{equation}
    <NMI> = \frac{2}{K(K-1)} \sum_{a=1}^{K-1}\sum_{b=a+1}^{K} NMI(p_a,p_b)
\end{equation}

where $NMI(p_a, p_b)$ is the Normalized Mutual Information between a given pair of partitions, and $\frac{2}{K(K-1)}$ is the number of distinct pairs. The majority partition over K = 100 runs can be approved only if $\langle \text{NMI} \rangle$ $\geq$ 0.90, and if it is non-trivial (i.e., not consisting of a single community) and if it is not too fragmented, namely it does not contain communities whose cardinality is less than 5$\%$ of the cardinality of the partitioned network.

If the majority voting results obtained for 100 runs, at different values of the resolution $\gamma$, satisfy the above conditions, we choose the output with larger $\langle \text{NMI} \rangle$, and the majority partition corresponding to this choice is identified as the result of community detection. The procedure is applied in a hierarchical algorithm, in which at each step the communities obtained at the previous step are partitioned according to the same criteria. The hierarchical community detection stops when no partition obtained at a given step satisfies the stability, non-triviality and non-fragmentation criteria.

\subsection{Latent Ideology Estimation}
To estimate the internal organization of the coalitions, we employ a correspondence analysis based model. This model is originally proposed to study ideological positions in social networks based on homophily \cite{barbera2015birds, barbera2015tweeting}, and it is subsequently adapted to quantify polarization in online environments \cite{flamino2023political, falkenberg2022growing}.

In this study, we transpose this approach to the electoral domain to investigate the intra-coalition dynamics. A key theoretical challenge arises when adapting polarization metrics, typically designed for binary oppositions (e.g.: Left vs. Right), to a multi-candidate coalition. We consider that due to the specific electoral law, which limits voters to expressing at most two preferences, induces an \emph{emergent binarization}. In fact, the scarcity of expressible preferences forces a selective choice, compelling candidates to organize into \emph{de facto} subgroups (or ``tickets'') to maximize consensus. Consequently, even within a single coalition, voting patterns are rarely uniform. By applying latent-ideology theory at the intra-coalition level, we aim to assess whether such selective dynamics become accentuated: ideally, if the votes recorded within a given polling section were concentrated on two candidates, this would correspond to a high latent-ideology value, since it reflects a polarized choice centered on two alternatives. Conversely, a low latent-ideology value supports the presence of a more homogeneous vote distribution within that polling section.

In order to study the organization into coalitions, our goal is thus to identify the latent structure for each of them. To formulate the analysis at the level of individual coalition networks, we move beyond simple vote counts to identify the direction of preference: distinguishing as $monoliths$ (i.e. coalition supporting all candidates in homogeneous-like way) from $polarized$ (i.e. coalition supporting specific internal candidates to the exclusion of others).

Let the voting data for a specific coalition be represented by an interaction matrix
\[
A = (A_{ij})_{i=1,\dots,n}^{j=1,\dots,m},
\]
where $A_{ij} \in \mathbb{R}_{\ge 0}$ denotes the number of votes that polling section $i$ assigns to candidate $j$.
Define the total number of observed votes as
\[
N = \sum_{i=1}^n \sum_{j=1}^m A_{ij},
\]
and normalize the matrix to obtain the empirical joint distribution
\[
P_{ij} = \frac{A_{ij}}{N},
\qquad \sum_{i,j} P_{ij} = 1.
\]

The row and column marginals,
\[
r_i = \sum_{j=1}^m P_{ij}, \qquad
c_j = \sum_{i=1}^n P_{ij},
\]
encode, respectively, the relative weight of each section and the aggregate popularity of each candidate. Under an independence model (where every coalition supports candidates strictly according to their global popularity), the expected joint distribution would be
\[
P^{\text{ind}}_{ij} = r_i c_j.
\]

To evaluate deviations from this independence, which reveal the specific territorial affinities, we introduce the diagonal matrices
\[
D_r = \mathrm{diag}(r_1, \dots, r_n),
\qquad
D_c = \mathrm{diag}(c_1, \dots, c_m),
\]
and construct the standardized residual matrix
\[
S = D_r^{-1/2} \, (P - r c^\top) \, D_c^{-1/2},
\]
whose entries are explicitly given by
\[
S_{ij} = \frac{P_{ij} - r_i c_j}{\sqrt{r_i c_j}}.
\]

The matrix $S$ captures the association structure after removing marginal effects. We then perform the singular value decomposition (SVD) to extract the principal dimensions of variance:
\[
S = U \Sigma V^\top,
\]
where $U$ and $V$ are orthonormal and
\[
\Sigma = \mathrm{diag}(\sigma_1, \dots, \sigma_k),
\qquad k = \min(n,m),
\]
contains the singular values in decreasing order.

The first left singular vector,
\[
u^{(1)} = U_{:,1},
\]
identifies the direction in the space of polling sections that maximizes the explained association with the voting data. In our intra-coalition context, $u^{(1)}$ represents the axis of emergent binarization. It separates the precincts based on the primary internal cleavage of the coalition.

To obtain stable and comparable ideology scores, we standardize this vector:
\[
\text{Ideology}_i
=
\frac{u^{(1)}_i - \mu}{\sigma},
\]
where $\mu$ and $\sigma$ denote the sample mean and standard deviation of $u^{(1)}$.

The resulting values represent the latent ideological coordinates. It is important to emphasize that in this restricted scenario, these coordinates do not map the global Left-Right spectrum \cite{di2024sampled}. Rather, they measure the alignment of a coalition with the distinct internal currents of the coalition.
\begin{itemize}
    \item A value near zero indicates a coalitions where the voting distribution mirrors the coalition's average ($monoliths$, territorial homogeneity).
    \item High absolute values (positive or negative) indicate coalitions that disproportionately favor a specific subgroup of candidates over others ($polarized$, internal polarization).
\end{itemize}
This metric, therefore, serves as a quantitative proxy for the structural solidity or fragmentation of the consensus across the city.

\subsection{Analysis of electoral fragmentation through Shannon Entropy}

The distribution of latent ideology values $\{I_1, I_2, \dots, I_N\}$ is observed for each coalition across 345 electoral sections through the calculation of first- and second-order moments. We define the sample mean $\hat{\mu}$ and the sample variance $\hat{\sigma}^2$ as fundamental indicators of the central tendency and consensus dispersion for each coalition:

\begin{equation}
\hat{\mu} = \frac{1}{N} \sum_{i=1}^{N} x_i, \quad \hat{\sigma}^2 = \frac{1}{N-1} \sum_{i=1}^{N} (x_i - \hat{\mu})^2
\end{equation}

Although these parameters offer a preliminary description of the ideological center of gravity, their validity as exhaustive descriptors strictly depends on the assumption of data normality. To verify this hypothesis, it is customary to subject the distributions to the Shapiro-Wilk test \cite{gonzalez2019shapiro}, which evaluates the null hypothesis $H_0$ that the samples originate from a Gaussian-distributed population. If the Shapiro-Wilk test rejects this hypothesis ($p \ll 0.05$) and a high distribution asymmetry (measured via skewness, that is 0 for ideal Gaussian distribution) and excessive kurtosis (measured via leptokurtosis, which have a kurtosis \(>3\), whereas Gaussian distributions have a Kurtosis \(=3\)) are detected, this symbolizes the presence of heavy tails and multimodal structures. These phenomena, which are interpretable as signals of the $binarization$ of the latent ideology distribution and thus territorial fractures, render the Gaussian approximation insufficient to describe the complexity of the electoral reality.

As it is possible to see in $Results$ section, to overcome the limitations of parametric models, it is possible to analyse those distribution by mean of Shannon Entropy ($H$) \cite{lin2002divergence}, i.e. as a rigorous and non-parametric measure of fragmentation \cite{marmani2020entropic}. This approach, borrowed from statistical physics, allows us to quantify the degree of disorder or heterogeneity of the system without imposing a predefined functional form on the distribution. To proceed with the calculation of $H$, we discretize the continuous space of latent ideology into $K$ intervals of equal width. We thus define the probability $p_k$ that the ideology value calculated for an electoral precinct falls into the $k$-th bin as:

\begin{equation}
p_k = \frac{n_k}{N}
\end{equation}

where $n_k$ is the observed frequency in the bin. Shannon Entropy is then calculated as:

\begin{equation}
H(X) = - \sum_{k=1}^{K} p_k \log_2(p_k)
\end{equation}

In our framework, we interpret $H$ as a direct proxy for territorial fragmentation: an $H$ value close to zero indicates maximum homogeneity and consensus cohesion, while high values signal a heterogeneous distribution of votes across different ideological spectra (see $Results$ for more details). Since the latent ideology value is a continuous variable, the choice of a discretization interval is necessary. Since each coalition comprises 345 electoral sections, a value of $K$=30 provides a robust resolution for estimating electoral patterns. This choice ensures the statistical significance of each bin and exceeds the requirements of Sturges' rule \cite{scott2009sturges} to better capture the potential bimodality and heavy tails observed in the latent ideology distributions. In fact, since each ideology distribution has a number of observables $N=345$, a value of $K=30$ allows for an average of approximately 11 values per bin. This is a sufficiently number to ensure that the probabilities $p_k$ are statistically significant and not random fluctuations of a single precinct, in order to allow secondary peaks (binarization) to emerge while maintaining a high enough number of observations per bin to avoid over-fitting.

Since $H$ is a non-linear function of sample probabilities, it is highly sensitive to the stochastic fluctuations of the original dataset and tends to exhibit a systematic bias that can lead to an underestimation of the true informational disorder \cite{bonachela2008entropy}. In the absence of a known theoretical distribution for the latent ideology values, due to a possible non-Gaussian nature of the data, it is not possible to apply error propagation formulas based on Gaussian statistics. Consequently, to estimate the uncertainty associated with entropy $H$, we apply a bootstrap procedure, generating 1,000 sub-samples through resampling with replacement from the original dataset.

\section{Results}

\subsection{Representing electoral section by means of complex network}
\begin{figure*}[t]
    \centering
    \includegraphics[width=0.85\textwidth]{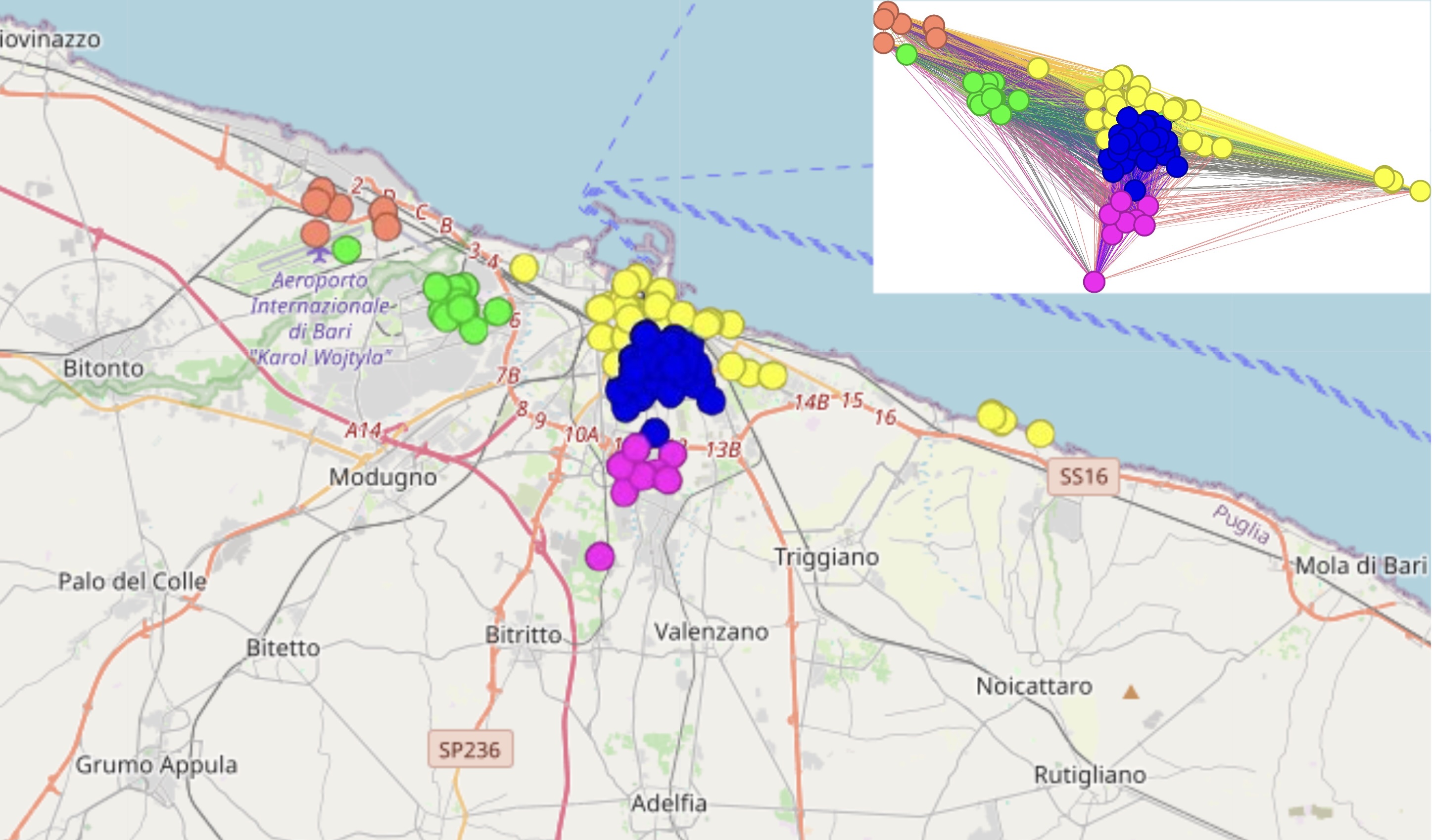}
\caption{Representation of polling sections through a complex network. In main panel, we present the electoral sections colored by administrative membership division of the City of Bari into its five municipalities. The upper-right panel illustrates the geographical distribution of the 345 polling sections using a complex network. In this network, each node represents a polling section, and each link is a statistically significant correlation between two nodes (if present). Here, yellow nodes refers to Municipality 1 (M1) which includes the quarters of Murat, San Nicola, Libertà, Madonnella, Japigia, Torre a Mare. Blue nodes refers to Municipality 2 (M2), gathering Picone, Poggiofranco, Carrassi, San Pasquale and Mungivacca quarters. Green nodes includes Marconi, San Girolamo, Fesca, San Paolo, Stanic and Villaggio del Lavoratore quartes, indicating the Municipality 3 (M3) while Carbonara di Bari, Ceglie del Campo and Loseto are included in Municipality 4 (M4). Finally the Municipality 5 (M5) includes the quartes of Palese, Macchie, Santo Spirito, Catino and San Pio. All the administrative information are extracted from the City of Bari official website \cite{bari_municipi}. Network representation is constructed using Gephi software with GeoLayout (Licence GPL-3.0). See administrative municipality membership in the $Supplementary$ $Information$ for more details. \cite{lo_sasso_dataset_election}}
\label{fig:municipi}
\end{figure*}

As detailed in $Methods$, to develop a model capable of describing the electoral votes at local level, and evaluate the behaviour recorded in electoral section by means of correlations, we constructed a weighted complex network of polling stations. In this network, each node is a section, and link between two nodes is present if there is a statistically significant correlation exists among them. We have globally 345 nodes, one for each electoral section. We calculate the statistically significant correlation using the Pearson correlation of the vote vectors recorded by the two sections ($p_{val} < 10^{-6}$). The weight of link is the correlation value we assign to the link among two electoral section. Figure \ref{fig:municipi} illustrates the territorial distribution of the polling sections and their representation as a complex network, where electoral sections are coloured according to the administrative municipal membership, and they are also located based on the geographical coordinates of the section. All information regarding polling section, as well as location and administrative municipality membership, are provided in the $Supplementary$ $Information$ \cite{lo_sasso_dataset_election}.

In order to describe the voting behaviour of the electoral coalitions separately, we construct one network for each coalition. As detailed in $Methods$, we analyze $C_{1}^{M}$ (centre-left coalition), $C_{2}^{M}$ (right coalition) and $C_{3}^{M}$, (left coalition) for mayoral election, while $C_{1}^{R}$ (centre-left coalition) and ($C_{2}^{R}$) (centre-right coalition) for regional one.

For networks describing \(C_{1}^{M}\), \(C_{2}^{M}\), and \(C_{3}^{M}\), the nodes represent the polling stations, and the weighted link between two polling stations is evaluated considering the statistically significant Pearson correlation basing on recorded vote solely by each coalition, separately. The same procedure was applied to \(C_{1}^{R}\), \(C_{2}^{R}\), where each correlation for each coalition is computed solely on the electoral outcomes that the coalition obtains across the 345 sections in Bari area.

To identify groups of polling sections that are more strongly correlated based on expressed electoral preferences, we conduct a community detection analysis to examine coalition behaviour. This approach helps to reveal elective affinities among the nodes of the network. In order to have a descriptive partition of the nodes of each network, we thus employ the Leiden algorithm \cite{traag2019louvain}, optimizing modularity in extracting communities from networks. The criteria and parameters we adopt in community detection are described in Methods. We use the Normalized Mutual Information (NMI) to evaluate the goodness of subdivision, repeating partition 100 times, and changing the random seed at each repetition. An NMI value close to 1 indicates that, across repetitions, the partitions tend to yield the same communities, whereas an NMI value close to 0 reflects communities whose composition varies substantially at each iteration.

For the three coalition in mayoral election, the network associated with $C_{1}^{M}$ was divided into three communities, containing 143 nodes, 165 nodes, and 37 nodes, respectively (NMI = 1). Since the first two $C_{1}^{M}$ partition meet the conditions for a further community partitioning (see Method section for more details), a second subdivision of the first two communities is performed. The Leiden algorithm divides the community with 143 nodes into three sub-communities of 47, 64, and 32 nodes, while the community with 165 nodes was split into three sub-communities of 58, 50, and 57 nodes (NMI = 0.99 and NMI = 0.96 respectively). The community of 37 nodes is not further subdivided. For coalition $C_{2}^{M}$, the network was initially divided into 6 communities, containing 114 nodes, 82 nodes, 41 nodes, 106 nodes, and 2 communities with a single node each, with an obtained NMI equals to 0.83. We stop here to perform further subdivisions of the communities in $C_{2}^{M}$ because the conditions of homogeneity and non-triviality useful for another subdivision are not meet. Thus, the network associated to coalition $C_{3}^{M}$ is divided into 5 communities, with sizes of 159 nodes, 106 nodes, 35 nodes, 44 nodes, and 1 community with a single node (NMI = 0.99). Similar to $C_{2}^{M}$, no further subdivisions of the communities in $C_{3}$ are performed as the conditions for subsequent division are missing.

For what concern the two coalition in regional election, the network associated with $C_{1}^{R}$ is divided into three communities, containing 148 nodes, 106 nodes, and 91 nodes, respectively (NMI = 0.98). Since the first two $C_{1}^{M}$ partition meets the conditions for a further community partitioning, a second subdivision of the first two communities is performed. We further subdivided these larger communities, containing 85 and 65 for first sub-partition, while 23 and 83 for the second one (NMI = 0.96). At the same time $C_{2}^{R}$ is divided into three communities, containing 154 nodes, 59 nodes, and 132 nodes, respectively (NMI = 0.91). Since the first and the third $C_{2}^{M}$ partition meets the conditions for a further community partitioning, a second subdivision of the those communities is performed. We further subdivided these communities, containing 83 and 71 for first sub-partition and 93 and 39 for the last one.

\begin{figure*}[t]
    \centering
    \includegraphics[width=0.85\textwidth]{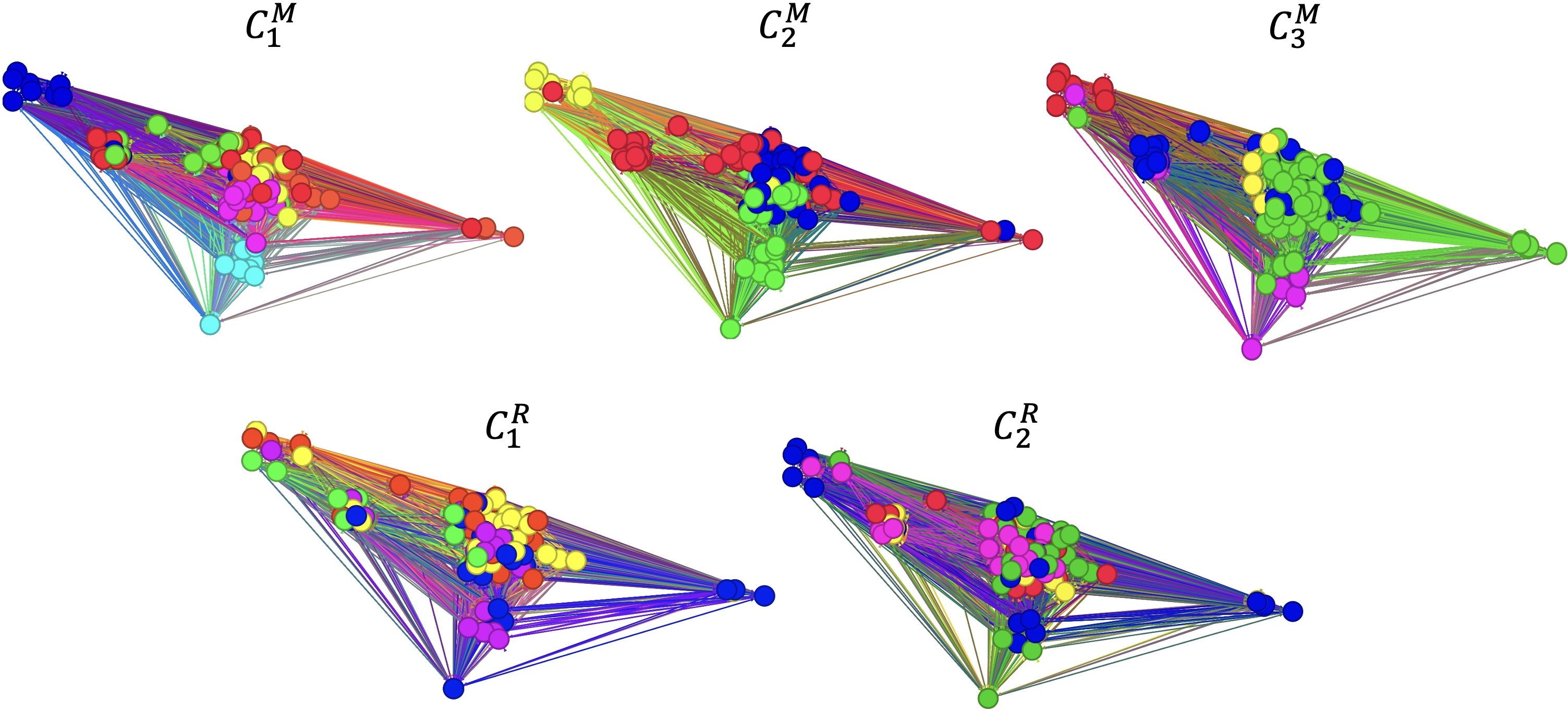}
\caption{Community membership of polling sections. Each node in the network, for each coalition, is coloured according to its community assignment. The analysis shows that, in the city of Bari, the geographical distribution of communities within the coalition networks for the mayoral and regional elections reveals that each coalition possesses sections with shared political affinities, which are also geographically close to one another.}
\label{fig:network_coalitions}
\end{figure*}

In Figures \ref{fig:network_coalitions}, we illustrate the geographically located electoral sections for the three coalition network, coloured now according to community membership for each coalition. We note that electoral sections exhibiting similar voting behaviour (i.e. presenting the same color in community membership) also exhibit geographical proximity. This geographical pattern is more pronounced for coalitions $C_{2}^{M}$ and $C_{3}^{M}$, whereas the communities appear to be more locally dispersed across the territory in the case of coalitions $C_{1}^{M}$, $C_{1}^{R}$, and $C_{2}^{R}$.

Referring to Figure \ref{fig:network_coalitions} and juxtaposing it with the administrative subdivision of polling sections reported in Figure \ref{fig:municipi}, several cross-municipality affinity patterns become apparent. For instance, coalition $C_2^{M}$ exhibits closely aligned voting profiles in both M4 and M2 (the community highlighted by green nodes), while a distinct community associated with $C_2^{M}$ displays comparable behaviour across M3 and selected sections of M1. Similarly, in the regional election, coalition $C_2^{R}$—which, as detailed in the Methods, pertains to the same political faction—shows analogous electoral affinities between M1 and M2. Turning to the centre-left side (represented by two coalitions at the municipal level and one at the regional level), informative structures also emerge for $C_1^{M}$, $C_3^{M}$, and $C_1^{R}$. In particular, $C_3^{M}$ forms a cohesive block of mutually similar sections spanning M1, M2, and M4, whereas $C_1^{M}$ yields a pronounced aggregate encompassing M5, M4, and M2. In the regional contest, $C_1^{R}$ instead appears comparatively compact, with a clustered pattern concentrated in M1 and M5.

\subsection{Latent ideology estimation and Shannon entropy as a measure of fragmentation}
Community detection highlights the presence of territorial structure in voting behaviour. However, community detection alone is insufficient to identify the degree of fragmentation or order within coalition behavior. To measure the intensity of the emergent behavioral structure, we introduce an intra-coalition latent-ideology analysis, the mathematical formalism of which is detailed in the $Methods$ section. For each node in the networks of every coalition, we calculated a latent ideology value and examined its whole coalition distribution. While classical literature typically assumes that such distributions can be approximated by a Gaussian distribution \cite{kittler2003sum}, our findings suggest otherwise. In fact, we verify whether the coalition distributions follow a Gaussian functional form by evaluating this behaviour through statistical testing on the empirical data. Table \ref{tab:normality_test} exhibits structural anomalies that invalidate this assumption, by adoption the Shapiro-Wilk test \cite{gonzalez2019shapiro}. Despite the calculation of first and second-order moments (namely the mean $\mu$ and standard deviation $\sigma$), statistical inference tests consistently rejected the null hypothesis of normality. In fact for all analyzed coalitions, the third and fourth moment of distribution do not follow the Gaussian shape values (see Methods section for more detail), and the p-values were significantly below the critical threshold (p $\ll$ 0.05 according to the Shapiro-Wilk test \cite{gonzalez2019shapiro}), indicating that the probability of the data originating from a normal population is statistically negligible.

\begin{table*}[t]
\centering
\begin{tabular}{@{}lccccc@{}}
\toprule
\textbf{Coalition} & \textbf{Mean $\mu$} & \textbf{Std. Dev. $\sigma$} & \textbf{Skewness} & \textbf{Kurtosis} & \textbf{Shapiro $p$-value} \\ \midrule
$C_1^{M}$ & 0.030 & 0.165 & 1.10 & 5.26 & $6.28 \times 10^{-12}$ \\
$C_2^{M}$ & -0.234 & 0.263 & 5.90 & 50.74 & $4.16 \times 10^{-29}$ \\
$C_3^{M}$ & 0.307 & 0.234 & -3.19 & 26.60 & $1.20 \times 10^{-21}$ \\
$C_1^{R}$ & 0.337 & 0.187 & -0.42 & 1.86 & $3.12 \times 10^{-4}$ \\
$C_2^{R}$ & -0.215 & 0.664 & -0.75 & 6.01 & $2.06 \times 10^{-9}$ \\ \bottomrule
\end{tabular}
\caption{Fit validation of the Gaussian model approximation for each latent ideology distribution. Following the Shapiro-Wilk test, the hypothesis of Gaussian distribution is rejected.}
\label{tab:normality_test}
\end{table*}

The high kurtosis (leptokurtosis) and p-values close to zero in Table \ref{tab:normality_test} indicate the presence of heavy tails and systematic outliers for distribution of $C_1^{M}$, $C_2^{M}$, $C_3^{M}$, $C_1^{R}$ and $C_2^{R}$. Such evidence suggests that latent ideology is not centred around a single value but is instead structured into distinct blocks, making the Gaussian fit an incomplete description of electoral reality.

To obtain a quantitative measure that overcomes the limitations of parametric models (such as the Gaussian approximation), and to provide an accurate characterization of the latent ideology distribution within each coalition, we adopt Shannon Entropy $H$, which is widely used in statistical physics to measures the degree of fragmentation or heterogeneity of a system without assumptions regarding the distribution's shape. As we explain in $Methods$ section, the Shannon entropy quantifies territorial fragmentation for each coalition, where a high $H$ value indicates that a coalition draws votes from a diverse and non-homogeneous ideological spectrum, while $H$ close to 0 indicates and homogenous value indicates. In Table \ref{tab:entropy_results}, we report the $H$ value recorded for each coalition, together with the associated uncertainty estimated via a bootstrap procedure. Specifically, we generate 1000 bootstrap replicates by resampling the original dataset with replacement (see $Methods$ section for more detail).

\begin{table}[t]
\centering
\begin{tabular}{@{}lc@{}}
\toprule
\textbf{Coalition} & \textbf{Shannon Entropy ($H$)}\\ \midrule
$H_{C_1^{M}}$& 3.70 $\pm$ 0.11 \\
$H_{C_2^{M}}$& 2.38 $\pm$ 0.25 \\
$H_{C_3^{M}}$& 2.88 $\pm$ 0.30 \\
$H_{C_1^{R}}$& 3.78 $\pm$ 0.24 \\
$H_{C_2^{R}}$& 3.58 $\pm$ 0.17 \\ \bottomrule
\end{tabular}
\caption{Shannon Entropy $H$ of each coalition at fixed bin, $K=30$ bins (see $Methods$ for more details). To estimate the uncertainty associated with entropy $H$, we apply a bootstrap procedure, generating 1,000 sub-samples through resampling with replacement from the original dataset.}
\label{tab:entropy_results}
\end{table}

The entropy values presented in Table \ref{tab:entropy_results} demonstrate that, within the municipal elections, coalition $C_1^{M}$ is distinguished by a higher degree of fragmentation compared to the other competing coalitions. Specifically, we recorded $H_{C_1^{M}} = 3.70 \pm 0.11$, in contrast to $H_{C_2^{M}} = 2.38 \pm 0.25$ and $H_{C_3^{M}} = 2.88 \pm 0.30$. Similarly, transitioning to the regional elections, the entropy value for $C_1^{R}$ is $3.78 \pm 0.24$, while for $C_2^{R}$ we obtained a value of $3.58 \pm 0.17$. Shannon entropy value in Table \ref{tab:entropy_results} leads us to conclude that the winning coalitions in both electoral contests ($C_1^{M}$ and $C_1^{R}$, respectively) are characterized by higher entropy values relative to the other coalitions in the same competition, that means they are characterized by drawing votes from a diverse and non-homogeneous ideological spectrum.

\section{Discussion}
The study carried out on Bari polling sections, made comparable through the construction of complex networks based on correlations between the votes cast in each section, makes it possible to interpret local political affinities not only in aggregate terms, but also through a systematic reading of the internal dynamics of each coalition.

The construction of the complex network reveals the presence of coherent geographical community structures. In fact, juxlapping electoral section in Figure \ref{fig:municipi} and the community detection membership in Figure \ref{fig:network_coalitions}, it shoes that sections exhibiting similar electoral behaviour also tend to be geographically close to one another. The robustness of the partitions emerging from the community structure, obtained through the Leiden algorithm and evaluated by means of the NMI, indicates that these communities are not numerical artefacts, but stable structures reflecting territorially rooted preferences and urban areas characterised by analogous voting patterns.

Community detection analysis at the local level highlights that geographically proximal electoral sections also exhibit strong affinities in voting behaviour. This approach, quantitatively formulated through statistical physics techniques and complex networks, aligns with established evidence in the existing literature \cite{foladare1968effect}. In addition, Figure \ref{fig:network_coalitions} elucidates the differences between network nodes for each coalition. By integrating these findings with the territorial subdivision presented in Figure \ref{fig:municipi}, these data become instrumental for electoral strategies; in fact, they emphasize that diverse policy approaches are required even within the same city. Such strategies should not be strictly confined to administrative boundaries but should extend across broader, functionally connected territories.

Referring to Figure \ref{fig:network_coalitions} and comparing it with the administrative partition of polling sections shown in Figure \ref{fig:municipi}, we observe, for instance, that votes for coalition $C_2^{M}$ express similar preferences both in M4 and M2 (the community marked by green nodes), and that another $C_2^{M}$ community displays analogous behaviour across M3 and portions of M1. Moreover, regarding the regional elections into Bari section, the behaviour of coalition $C_2^{R}$ (which, as explained in the Methods section, belongs to the same right political faction) exhibits similar electoral affinities in M1 and M2. As for the centre-left coalitions (two coalitions in the municipal elections $C_1^{M}$, $C_3^{M}$ and $C_1^{R}$ in the regional contest), they nonetheless generate patterns of interest. For example, coalition $C_3^{M}$ shows an entire block of mutually similar voting sections spanning M1, M2, and M4, whereas $C_1^{M}$ presents a robust aggregate across M5, M4, and M2. On the other hand, in the regional election $C_1^{R}$ displays a compact behaviour in M1 and M5.

Although community detection identifies similarities among network nodes in the absence of an explicit spatial geometry, the results in Figure \ref{fig:network_coalitions} indicate that the emergent node similarity within these patterns also corresponds to geographical proximity. This insight previously established for national elections \cite{el2023entropic}, it is also manifested here at a local level, although it has never been analysed previously due to the profound complexity and variability of voting patterns at the local scale now detectable by means a complex network approach. Nevertheless, considered in isolation, these qualitative observations are insufficient to fully elucidate coalition fragmentation, as they do not provide a quantitative estimate of how homogeneous (or, conversely, how internally heterogeneous) each coalition is.

To elucidate coalition behaviour and thus derive insights for optimal electoral strategies, we employ the analysis of latent ideology. By performing a spectral analysis of the voting matrix, we identify the direction of maximum variance, which represents the primary ideological cleavage dividing the electorate. Even in races with multiple candidates, this method reveals whether precincts exhibit polarized voting outcomes, effectively bifurcated into opposing blocs, within a fixed coalition.

Since these ideological markers are derived from the preferences expressed by voters within each coalition, the measure reliably distinguishes whether a polling section exhibits a more polarized voting pattern, relative to that observed in other sections, or conversely, a more homogeneous one.

As illustrated in Table \ref{tab:normality_test}, the pronounced leptokurtosis and the near-zero p-values signal the existence of heavy tails and systematic outliers for each latent idealogy distribution. This evidence implies that latent ideology distribution of each coalition does not converge around a single ideological value, but rather, it is organized into distinct, localized clusters. Consequently, a standard Gaussian approximation fails to capture the inherent complexity of the electoral landscape.

To address the shortcomings of parametric modelling, we calculate the Shannon Entropy $H$, which serves as a measure of territorial fragmentation for each coalition, where a high H value reflects a coalition that aggregates votes from a diverse and non-homogeneous ideological spectrum, whereas an H value approaching zero indicates a high degree of ideological homogeneity.

The analysis of H values demonstrates that the winning coalitions consistently display higher internal fragmentation. This pattern is confirmed by $C_1^{M}$ in the municipal elections, which yielded $H_{C_1^{M}} = 3.70 \pm 0.11$, in contrast to $C_2^{M}$ and $C_3^{M}$, which recorded $H_{C_2^{M}} = 2.38 \pm 0.25$ and $H_{C_3^{M}} = 2.88 \pm 0.30$, respectively.

Globally, the entropy values align with studies analysing national-level elections \cite{el2023entropic, bailey2025politics}; however, to the best of our knowledge, no prior research has specifically examined individual coalitions. The higher H values linked to winning coalitions suggest that stratifying a coalition into multiple internal blocs is a successful strategy in both municipal and regional contests. Thus, these findings offer a significant implications for electoral strategy and ticket construction, as they indicate that selecting candidates capable of mobilizing localized voting blocs is more effective than forming coalitions dominated by a leading candidate who draws the majority of the coalition's votes, and so polarizing electorate few people only.

While this study rigorously applies statistical-physics techniques to extend quantitative voting analysis from the national to the local level—thereby opening the possibility of studying elections at the local scale through a relational, complex-network-based approach that can effectively handle and describe highly complex phenomena, this framework still exhibits several limitations that should serve as points of reflection and motivation for future work.

First, the analysis is restricted to the city of Bari. This constraint is primarily due to the scarcity of available data on the $Eligendo$ platform and from other municipal statistical offices. Since granular information at the polling station level is only accessible on $Eligendo$ for a limited period following an election, it was not possible to reconstruct the electoral behaviour of other cities involved in the 2024 cycle, nor to extend the analysis to the 2025 regional elections in other areas, including the broader Apulia region. This line of inquiry remains a priority for future research.

Furthermore, regarding the analysis of electoral and territorial affinities, this study relies on a descriptive community detection approach. As such, it provides a snapshot of the qualitative structure of electoral partitions post-hoc, without investigating the underlying causal mechanisms driving these preferences. Future studies should, therefore, incorporate electoral choices in relation to the socio-economic status of voters. Indeed, the literature indicates that divergent socio-economic conditions significantly influence alignment with left- or right-wing parties \cite{lois2025effects, leigh2005economic}, a phenomenon particularly documented within the Italian context \cite{bellucci1984effect}. Integrating such variables would help account for the observed heterogeneities in voting behaviour, linking them to the relative wealth and living conditions of residents across different urban neighborhoods.

\printbibliography

@article{hidalgo2007product,
  title={The product space conditions the development of nations},
  author={Hidalgo, C{\'e}sar A and Klinger, Bailey and Barab{\'a}si, A-L and Hausmann, Ricardo},
  journal={Science},
  volume={317},
  number={5837},
  pages={482--487},
  year={2007},
  publisher={American Association for the Advancement of Science}
}

@article{battiston2012debtrank,
  title={Debtrank: Too central to fail? financial networks, the fed and systemic risk},
  author={Battiston, Stefano and Puliga, Michelangelo and Kaushik, Rahul and Tasca, Paolo and Caldarelli, Guido},
  journal={Scientific reports},
  volume={2},
  number={1},
  pages={1--6},
  year={2012},
  publisher={Nature Publishing Group}
}

@article{bardoscia2021physics,
  title={The physics of financial networks},
  author={Bardoscia, Marco and Barucca, Paolo and Battiston, Stefano and Caccioli, Fabio and Cimini, Giulio and Garlaschelli, Diego and Saracco, Fabio and Squartini, Tiziano and Caldarelli, Guido},
  journal={Nature Reviews Physics},
  volume={3},
  number={7},
  pages={490--507},
  year={2021},
  publisher={Nature Publishing Group UK London}
}

@article{bellantuono2022territorial,
  title={Territorial bias in university rankings: a complex network approach},
  author={Bellantuono, Loredana and Monaco, Alfonso and Amoroso, Nicola and Aquaro, Vincenzo and Bardoscia, Marco and Loiotile, Annamaria Demarinis and Lombardi, Angela and Tangaro, Sabina and Bellotti, Roberto},
  journal={Scientific reports},
  volume={12},
  number={1},
  pages={4995},
  year={2022},
  publisher={Nature Publishing Group UK London}
}

@article{bellantuono2025network,
  title={Network assortativity for a multidimensional evaluation of socio-economic territorial biases in university rankings},
  author={Bellantuono, Loredana and Lo Sasso, Andrea and Amoroso, Nicola and Monaco, Alfonso and Tangaro, Sabina and Bellotti, Roberto},
  journal={PLoS One},
  volume={20},
  number={6},
  pages={e0323356},
  year={2025},
  publisher={Public Library of Science San Francisco, CA USA}
}

@article{amoroso2018multiplex,
  title={Multiplex networks for early diagnosis of Alzheimer's disease},
  author={Amoroso, Nicola and La Rocca, Marianna and Bruno, Stefania and Maggipinto, Tommaso and Monaco, Alfonso and Bellotti, Roberto and Tangaro, Sabina},
  journal={Frontiers in Aging Neuroscience},
  volume={10},
  pages={365},
  year={2018},
  publisher={Frontiers Media SA}
}

@article{bellantuono2021predicting,
  title={Predicting brain age with complex networks: From adolescence to adulthood},
  author={Bellantuono, Loredana and Marzano, Luca and La Rocca, Marianna and Duncan, Dominique and Lombardi, Angela and Maggipinto, Tommaso and Monaco, Alfonso and Tangaro, Sabina and Amoroso, Nicola and Bellotti, Roberto},
  journal={NeuroImage},
  volume={225},
  pages={117458},
  year={2021},
  publisher={Elsevier}
}

@incollection{batty2021london,
  title={London in lockdown: Mobility in the pandemic city},
  author={Batty, Michael and Murcio, Roberto and Iacopini, Iacopo and Vanhoof, Maarten and Milton, Richard},
  booktitle={COVID-19 Pandemic, Geospatial Information, and Community Resilience},
  pages={229--244},
  year={2021},
  publisher={CRC Press}
}

@article{foladare1968effect,
  title={The effect of neighborhood on voting behavior},
  author={Foladare, Irving S},
  journal={Political Science Quarterly},
  volume={83},
  number={4},
  pages={516--529},
  year={1968},
  publisher={JSTOR}
}

@article{lyra2003generalized,
  title={Generalized Zipf's law in proportional voting processes},
  author={Lyra, ML and Costa, UMS and Costa Filho, RN and Andrade Jr, JS},
  journal={Europhysics letters},
  volume={62},
  number={1},
  pages={131},
  year={2003},
  publisher={IOP Publishing}
}

@article{di2024sampled,
  title={Sampled Datasets Risk Substantial Bias in the Identification of Political Polarization on Social Media},
  author={Di Bona, Gabriele and Fraxanet, Emma and Komander, Bj{\"o}rn and Sasso, Andrea Lo and Morini, Virginia and Vendeville, Antoine and Falkenberg, Max and Galeazzi, Alessandro},
  journal={arXiv preprint arXiv:2406.19867},
  year={2024}
}

@misc{lo_sasso_dataset_election,
  title = {{Electoral results dataset -- Bari municipal (2024) and regional (2025) elections, by polling section}},
  author = {Lo Sasso, Andrea},
  howpublished = {\url{https://docs.google.com/spreadsheets/d/19SsRJghCtXw8Za_WGt93Au42IRxxM_0O/edit?usp=share_link&ouid=112996383204713701579&rtpof=true&sd=true}},
  note = {Accessed: 2026-09-15}
}

@misc{bari_municipi,
  title = {{The municipalities of Bari (IT)}},
  howpublished = {\url{https://www.comune.bari.it/i-municipi}},
  note = {Accessed: 2025-06-10}
}

@article{tufekci2014engineering,
  title={Engineering the public: Big data, surveillance and computational politics},
  author={Tufekci, Zeynep},
  journal={First Monday},
  year={2014}
}

@article{kaufman2022statistical,
  title={Statistical mechanics of political polarization},
  author={Kaufman, Miron and Kaufman, Sanda and Diep, Hung T},
  journal={Entropy},
  volume={24},
  number={9},
  pages={1262},
  year={2022},
  publisher={MDPI}
}

@article{gsanger2024opinion,
  title={Opinion models, election data, and political theory},
  author={Gs{\"a}nger, Matthias and H{\"o}sel, Volker and Mohamad-Klotzbach, Christoph and M{\"u}ller, Johannes},
  journal={Entropy},
  volume={26},
  number={3},
  pages={212},
  year={2024},
  publisher={MDPI}
}

@article{siegenfeld2020negative,
  title={Negative representation and instability in democratic elections},
  author={Siegenfeld, Alexander F and Bar-Yam, Yaneer},
  journal={Nature Physics},
  volume={16},
  number={2},
  pages={186--190},
  year={2020},
  publisher={Nature Publishing Group UK London}
}

@article{kaufman2024social,
  title={Social depolarization: Blume--capel model},
  author={Kaufman, Miron and Kaufman, Sanda and Diep, Hung T},
  journal={Physics},
  volume={6},
  number={1},
  pages={138--147},
  year={2024},
  publisher={MDPI}
}

@article{lo2025exploring,
  title={Exploring the relationship between cancer incidence and the sustainable development goals through complex networks and machine learning},
  author={Lo Sasso, Andrea and Bellantuono, Loredana and Omodei, Elisa},
  journal={EPJ Data Science},
  volume={14},
  number={1},
  pages={76},
  year={2025},
  publisher={Springer}
}

@article{borghesi2012election,
  title={Election turnout statistics in many countries: similarities, differences, and a diffusive field model for decision-making},
  author={Borghesi, Christian and Raynal, Jean-Claude and Bouchaud, Jean-Philippe},
  journal={PloS one},
  volume={7},
  number={5},
  pages={e36289},
  year={2012},
  publisher={Public Library of Science San Francisco, USA}
}

@article{auconi2024fluctuations,
  title={Fluctuations and extreme events in the public attention on Italian legislative elections},
  author={Auconi, Andrea and Federico, Lorenzo and Riotta, Gianni and Caldarelli, Guido},
  journal={Scientific Reports},
  volume={14},
  number={1},
  pages={22804},
  year={2024},
  publisher={Nature Publishing Group UK London}
}

@article{barbera2015birds,
  title={Birds of the same feather tweet together: Bayesian ideal point estimation using Twitter data},
  author={Barber{\'a}, Pablo},
  journal={Political analysis},
  volume={23},
  number={1},
  pages={76--91},
  year={2015},
  publisher={Cambridge University Press}
}

@article{barbera2015tweeting,
  title={Tweeting from left to right: Is online political communication more than an echo chamber?},
  author={Barber{\'a}, Pablo and Jost, John T and Nagler, Jonathan and Tucker, Joshua A and Bonneau, Richard},
  journal={Psychological science},
  volume={26},
  number={10},
  pages={1531--1542},
  year={2015},
  publisher={Sage Publications Sage CA: Los Angeles, CA}
}

@article{flamino2023political,
  title={Political polarization of news media and influencers on Twitter in the 2016 and 2020 US presidential elections},
  author={Flamino, James and Galeazzi, Alessandro and Feldman, Stuart and Macy, Michael W and Cross, Brendan and Zhou, Zhenkun and Serafino, Matteo and Bovet, Alexandre and Makse, Hern{\'a}n A and Szymanski, Boleslaw K},
  journal={Nature Human Behaviour},
  volume={7},
  number={6},
  pages={904--916},
  year={2023},
  publisher={Nature Publishing Group UK London}
}

@article{falkenberg2022growing,
  title={Growing polarization around climate change on social media},
  author={Falkenberg, Max and Galeazzi, Alessandro and Torricelli, Maddalena and Di Marco, Niccol{\`o} and Larosa, Francesca and Sas, Madalina and Mekacher, Amin and Pearce, Warren and Zollo, Fabiana and Quattrociocchi, Walter and others},
  journal={Nature Climate Change},
  volume={12},
  number={12},
  pages={1114--1121},
  year={2022},
  publisher={Nature Publishing Group UK London}
}

@article{traag2019louvain,
  title={From Louvain to Leiden: guaranteeing well-connected communities},
  author={Traag, Vincent A and Waltman, Ludo and Van Eck, Nees Jan},
  journal={Scientific reports},
  volume={9},
  number={1},
  pages={1--12},
  year={2019},
  publisher={Nature Publishing Group}
}

@article{hossain2017people,
  title={People’s voting behavior in local election: A study on Annadanagar Union Parishad, Pirgachha, Rangpur},
  author={Hossain, KR and Aktar, M and Islam, S},
  journal={IOSR Journal of Humanities and Social Science},
  volume={22},
  number={3},
  pages={1--14},
  year={2017}
}

@article{wildgen1980spatial,
  title={Spatial distribution of partisan support and the seats/votes relationship},
  author={Wildgen, John K and Engstrom, Richard L},
  journal={Legislative Studies Quarterly},
  pages={423--435},
  year={1980},
  publisher={JSTOR}
}

@article{barkan2006space,
  title={Space matters: Designing better electoral systems for emerging democracies},
  author={Barkan, Joel D and Densham, Paul J and Rushton, Gerard},
  journal={American Journal of Political Science},
  volume={50},
  number={4},
  pages={926--939},
  year={2006},
  publisher={Wiley Online Library}
}

@inproceedings{coscia2012towards,
  title={Towards democratic group detection in complex networks},
  author={Coscia, Michele and Giannotti, Fosca and Pedreschi, Dino},
  booktitle={International Conference on Social Computing, Behavioral-Cultural Modeling, and Prediction},
  pages={105--113},
  year={2012},
  organization={Springer}
}

@article{bates1996effects,
  title={The effects of sample size and variability on the correlation coefficient.},
  author={Bates, BARRY T and Zhang, SONGNING and Dufek, JANET S and Chen, FANG C},
  journal={Medicine and Science in Sports and Exercise},
  volume={28},
  number={3},
  pages={386--391},
  year={1996}
}

@article{foti2011nonparametric,
  title={Nonparametric sparsification of complex multiscale networks},
  author={Foti, Nicholas J and Hughes, James M and Rockmore, Daniel N},
  journal={PloS one},
  volume={6},
  number={2},
  pages={e16431},
  year={2011},
  publisher={Public Library of Science San Francisco, USA}
}

@article{monroe2001paradigm,
  title={Paradigm shift: from rational choice to perspective},
  author={Monroe, Kristen Renwick},
  journal={International Political Science Review},
  volume={22},
  number={2},
  pages={151--172},
  year={2001},
  publisher={Sage Publications Sage CA: Thousand Oaks, CA}
}

@article{lois2025effects,
  title={Effects of Socioeconomic Status on Right-Wing Voting Intentions: The Mediating Role of Economic Ideology, Perceived Threats, and National Identity},
  author={Lois, Giannis and Petkanopoulou, Katerina and Garc{\'\i}a-S{\'a}nchez, Efra{\'\i}n and Willis, Guillermo B and Rodr{\'\i}guez-Bail{\'o}n, Rosa},
  journal={Journal of Social and Political Psychology},
  volume={13},
  number={1},
  pages={80--99},
  year={2025}
}

@article{leigh2005economic,
  title={Economic voting and electoral behavior: How do individual, local, and national factors affect the partisan choice?},
  author={Leigh, Andrew},
  journal={Economics \& Politics},
  volume={17},
  number={2},
  pages={265--296},
  year={2005},
  publisher={Wiley Online Library}
}

@article{bellucci1984effect,
  title={The effect of aggregate economic conditions on the political preferences of the Italian electorate, 1953--1979},
  author={Bellucci, Paolo},
  journal={European Journal of Political Research},
  volume={12},
  number={4},
  pages={387--401},
  year={1984},
  publisher={Wiley Online Library}
}

@article{wagner2021affective,
  title={Affective polarization in multiparty systems},
  author={Wagner, Markus},
  journal={Electoral studies},
  volume={69},
  pages={102199},
  year={2021},
  publisher={Elsevier}
}

@article{kam2017polarization,
  title={Polarization in multiparty systems},
  author={Kam, Christopher and Indridason, Indridi and Bianco, William and others},
  journal={Polarization, institutional design and the future of representative democracy},
  year={2017}
}

@article{maza2022attempting,
  title={Attempting to measure the intensity of opposing feelings in elections: A polarization approach to Catalonia’s independence case},
  author={Maza, Adolfo and Hierro, Mar{\'\i}a},
  journal={Econom{\'\i}a Pol{\'\i}tica},
  volume={39},
  number={2},
  pages={323--344},
  year={2022},
  publisher={Springer}
}

@article{pal2025universal,
  title={Universal Statistics of Competition in Democratic Elections},
  author={Pal, Ritam and Kumar, Aanjaneya and Santhanam, MS},
  journal={Physical Review Letters},
  volume={134},
  number={1},
  pages={017401},
  year={2025},
  publisher={APS}
}

@article{perc2017statistical,
  title={Statistical physics of human cooperation},
  author={Perc, Matja{\v{z}} and Jordan, Jillian J and Rand, David G and Wang, Zhen and Boccaletti, Stefano and Szolnoki, Attila},
  journal={Physics Reports},
  volume={687},
  pages={1--51},
  year={2017},
  publisher={Elsevier}
}

@article{kittler2003sum,
  title={Sum versus vote fusion in multiple classifier systems},
  author={Kittler, Josef and Alkoot, Fuad M.},
  journal={IEEE transactions on pattern analysis and machine intelligence},
  volume={25},
  number={1},
  pages={110--115},
  year={2003},
  publisher={IEEE}
}

@article{gonzalez2019shapiro,
  title={Shapiro--Wilk test for skew normal distributions based on data transformations},
  author={Gonz{\'a}lez-Estrada, Elizabeth and Cosmes, Waldenia},
  journal={Journal of Statistical Computation and Simulation},
  volume={89},
  number={17},
  pages={3258--3272},
  year={2019},
  publisher={Taylor \& Francis}
}

@article{lin2002divergence,
  title={Divergence measures based on the Shannon entropy},
  author={Lin, Jianhua},
  journal={IEEE Transactions on Information theory},
  volume={37},
  number={1},
  pages={145--151},
  year={2002},
  publisher={IEEE}
}

@article{marmani2020entropic,
  title={Entropic analysis of votes expressed in Italian elections between 1948 and 2018},
  author={Marmani, Stefano and Ficcadenti, Valerio and Kaur, Parmjit and Dhesi, Gurjeet},
  journal={Entropy},
  volume={22},
  number={5},
  pages={523},
  year={2020},
  publisher={MDPI}
}

@article{scott2009sturges,
  title={Sturges' rule},
  author={Scott, David W},
  journal={Wiley Interdisciplinary Reviews: Computational Statistics},
  volume={1},
  number={3},
  pages={303--306},
  year={2009},
  publisher={Wiley Online Library}
}

@article{bonachela2008entropy,
  title={Entropy estimates of small data sets},
  author={Bonachela, Juan A and Hinrichsen, Haye and Munoz, Miguel A},
  journal={Journal of Physics A: Mathematical and Theoretical},
  volume={41},
  number={20},
  pages={202001},
  year={2008},
  publisher={IOP Publishing}
}

@article{el2023entropic,
  title={Entropic spatial auto-correlation of voter uncertainty and voter transitions in parliamentary elections},
  author={El Deeb, Omar},
  journal={Physica A: Statistical Mechanics and its Applications},
  volume={617},
  pages={128675},
  year={2023},
  publisher={Elsevier}
}

@article{bailey2025politics,
  title={Politics, Bit by Bit: A Formal Link Between Entropy and the Effective Number of Parties},
  author={Bailey, Jack},
  year={2025}
}



\end{document}